\documentclass[peerreview,11pt,draftcls,onecolumn]{IEEEtran}

\usepackage[T1]{fontenc}
\usepackage[latin9]{inputenc}
\usepackage{amsmath}
\usepackage{amssymb}
\usepackage{graphicx,color,url}

\usepackage{MnSymbol}
\def\x{{\mathbf x}}
\def\y{{\mathbf y}}
\def\k{{\mathbf k}}

\def\n{{\mathbf n}}

\def\z{{\mathbf z}}

\def\u{{\mathbf u}}

\def\D{{\cal D}}
\def\E{{\cal E}}

\def\K{{\cal K}}
\def\X{{\cal X}}

\def\DWR{{\mathsf{DWR}}}
\def\WNR{{\mathsf{WNR}}}
\def\SER{{\mathsf{SER}}}
\def\KK{{\mathbf K}}
\def\XX{{\mathbf X}}
\def\WW{{\mathbf W}}
\def\YY{{\mathbf Y}}
\def\NN{{\mathbf N}}
\def\OO{{\mathbf O}}
\def\UU{{\mathbf U}}
\def\set#1{\mathcal{#1}}
\def\Prob#1{\mathbb{P}\left[#1\right]}
\newcommand{\Exp}[2]{\mathbb{E}_{#1}[#2]}

\begin{document}

\title{The Effective Key Length of Watermarking Schemes}

\author{Patrick Bas and Teddy Furon%
\thanks{Patrick Bas is with CNRS-LAGIS, Ecole Centrale de Lille, Av. Paul Langevin, 59651 Villeneuve D'Ascq, France. \texttt{patrick.bas@ec-lille.fr}}%
\thanks{T. Furon is with Inria research centre Rennes Bretagne Atlantique, Campus de Beaulieu, 35042 Rennes, France.
\texttt{teddy.furon@inria.fr}}}

\markboth{IEEE Trans. on Information Forensics and Security,~Vol.~X, No.~Y, January~201Z}%
{Bas \& Furon: The Effective Key Length of Watermarking Schemes}

\maketitle
\begin{abstract}
Whereas the embedding distortion, the payload and the robustness of digital watermarking schemes are well understood, the notion of security is still not completely well defined. The approach proposed in the last five years is too theoretical and solely considers the embedding process, which is half of the watermarking scheme. This paper proposes a new measurement of watermarking security, called the \emph{effective key length}, which captures the difficulty for the adversary to get access to the watermarking channel. This new methodology is applied to additive spread spectrum schemes where theoretical and practical computations of the effective key length are proposed. It shows that these schemes are not secure as soon as the adversary gets observations in the Known Message Attack context.   
\end{abstract}

\begin{IEEEkeywords} Digital Watermarking, Security. \end{IEEEkeywords}

 \ifCLASSOPTIONpeerreview
 \begin{center} \bfseries EDICS Category: MOD-SECU, MOD-PERF, WAT-SSPM, WAT-THEO \end{center}
 \fi
%
\IEEEpeerreviewmaketitle

\section{Introduction}

\IEEEPARstart{F}{rom} the early beginning of its history, watermarking has been characterized by a trade-off between the embedding distortion and the capacity. The embedding distortion counts how hiding messages degrades the host contents. The capacity is the theoretical amount of hidden data that can be reliably transmitted when facing an attack of a given strength. In practice, the operating point of a watermarking technique is defined by the embedding distortion, the payload, and the robustness. These are well defined and gauged, for instance, by a Document to Watermarking power Ratio $\DWR$, a number of bits per host samples, and a Symbol Error Rate  $\SER$ at a given Watermark to Noise power Ratio $\WNR$.

Security came as a fourth feature stemming from applications where these exist attackers willing to circumvent watermarking such as copy and/or copyright protection. The efforts of the pioneering works introducing this new concept first focused on stressing the distinction between security and robustness. An early definition of security was coined by Ton Kalker as \emph{the inability by unauthorized users to have access to the raw watermarking channel}~\cite{KalkerSecurity}.     

The problem addressed in this paper is the following: the methodology to assess the security levels of watermarking schemes, proposed in~\cite{Cayre2005:Security,Comesana05,uvigo:tifs06security,gipsa:tifs07security,perez:tifs08security}, poorly captures T.~Kalker's definition. In a nutshell, the methodology proposed in these papers is based on C.~E.~Shannon's definition of security for symmetric crypto-systems~\cite{Shannon:1949:CSS}. The security level is defined as the amount of uncertainty the attacker has about the secret key. This is measured by the equivocation which is the entropy of the key knowing some observations such as contents watermarked with the same technique and the same secret key. 

Section~\ref{sub:PastApp} presents this past approach in more details and shows a surprising fact: this methodology only takes into account the embedding side. How could it capture the  `access to raw watermarking channel' in Kalker's definition if just half of the scheme is considered? 
Obviously, the decoding process should also play a role. Translating the theoretical foundations of cryptography security of~\cite{Shannon:1949:CSS} in watermarking terms may not have been a good idea. Indeed, watermarking and symmetric cryptography strongly disagree in the following point: In symmetric cryptography, the deciphering key is unique and is the ciphering key. Therefore, inferring this key from the observations (here, say some cipher texts) is the main task of the attacker. The disclosure of this key grants the adversary the access to the crypto-channel. In watermarking, several keys indeed can reliably decode hidden messages. Therefore, the precise disclosure of the secret key used at the embedding side is a possible way to get access to the watermarking channel, but it may not be the only one. 

As a solution, this article proposes an alternative methodology to assess the security level of a watermarking scheme as detailed in Sect.~\ref{sub:SecAlt}. In brief, our approach is based on the probability $P$ that the adversary finds a key that grants him the access to the watermarking channel as wished by Kalker: either a key decoding hidden messages embedded with the true secret key, either a key embedding messages that will be decoded with the true secret key. This gives birth to the concept of \emph{equivalent keys} presented in Sect.~\ref{sec:Definition-of-the}. Our new definition of the security level is called the \emph{effective key length} and is quantified by $\ell=-\log_{2}(P)$ in bits. This transposes the notion of cryptographic key length to watermarking: the bigger the effective key length, the smaller the probability of finding an equivalent key. This alternative methodology equally takes into account the embedding and the decoding sides. It is also simpler because it is not based on information theoretical notions and it allows to evaluate the effective key length experimentally (see Sect.~\ref{sec:Practical-computations}). 

The contributions of the paper are the following:
\begin{itemize}
\item A new methodology to estimate the security levels of watermarking schemes based on the definition of equivalent keys, the probability of finding such an equivalent key, and its translation in bits (Sect.~\ref{sec:Definition-of-the}).  
\item The application of this methodology to the Spread Spectrum (SS) watermarking scheme giving close form expressions of the  effective key length in Sect.~\ref{sec:Theoretical-computations}.
\item An experimental setup of Sect.~\ref{sec:Practical-computations} for estimating the effective key length with a comparison to the previous theoretical expressions.
\item The comparison of SS and ISS (Improved Spread Spectrum) watermarking techniques given in Sect.~\ref{sub:Experiments}.
\item The definitive evidence that these watermarking schemes have low security levels as soon as the adversary can get observations.
\end{itemize}

\section{Watermarking security}
This section details the methodology proposed so far to evaluate the security levels of watermarking schemes, and then it reviews our proposal.

\subsection{The past approach}\label{sub:PastApp}
We model the host by a vector $\x$ in set $\X$ extracted from a block of content.
Given a secret key $\k$, the embedding modifies this signal into vector $\y$ to hide message $m$: $\y=e(\x,m,\k)$.
The secret key is usually a signal: In spread spectrum schemes~\cite{CoxKilLeig}, the secret key is the set of carriers; in Quantization Index Modulation schemes~\cite{Chen:2001:QIM,EggersSCS}, it is the dither randomizing the quantization.
This signal is usually generated at the embedding and decoding sides thanks to a pseudo-random generated fed by a seed.
However, the attacker has no interest in disclosing this seed, because,
by analyzing watermarked contents, it is usually simpler to directly estimate $\k$ without knowing this seed.  

The attacker may disclose different kinds of information about the secret key.
First, he might get no information at all. This has been qualified as perfect covering in~\cite{Cayre2005:Security} or stego-security in~\cite{gipsa:tifs07security}. This happens when there is a total lack of identificability of the secret key.
A partial lack of identificability stems in different classes of security where the attacker only learns that the secret key lies in a given subset. For instance, in a spread spectrum scheme, he may learn that the watermark is added in a given subspace, however he may not identify the secret carriers up to a rotation matrix in this subspace. This is defined as subspace security in~\cite{gipsa:tifs07security}.

The application of the information theoretic approach of C.~E.~Shannon allowed to quantify watermarking security levels~\cite{Cayre2005:Security,perez:tifs08security,Comesana05,uvigo:tifs06security}.
This theory regards the signals used at the embedding as random variables (r.v.).
Let us denote $\KK$ the r.v. associated to the secret key, $\set{K}$ the space of the secret keys,
$\XX$ the r.v. associated to the host, $\set{X}$ the space of the hosts. 
Before producing any watermarked content, the designer draws the secret key $\k$ according to a given distribution $p_{\KK}$. The adversary knows $\set{K}$ and $p_{\KK}$ but he doesn't know the instantiation $\k$. This lack of knowledge is measured in bits by the entropy of the key $H(\KK)\triangleq-\sumint_{\set{K}}p_{\KK}(\k)\log_{2}p_{\KK}(\k)$ (i.e., an integral if $\KK$ is a continuous r.v. or a sum if $\KK$ is a discrete r.v.).

Now, suppose the adversary sees $N_{o}$ observations denoted as $\OO^{N_o}=\{\OO_{1},\ldots,\OO_{N_o}\}$.
The question is whether this key will remain a secret once the attacker gets these observations.
These include at least some watermarked contents which have been produced by the same embedder (same algorithm $e(\cdot)$, same secret key $\k$). These are also regarded as r.v. $\YY$. The observations may also encompass some other data depending on the attack setup (see definitions of WOA, KMA, KOA in~\cite{Cayre2005:Security}).

By carefully analyzing these observations, the attacker might deduce some information about the secret key.
The adversary can refine his knowledge about the key by constructing a posteriori distribution $p_{\KK}(\k|\OO^{N_o})$. The information leakage is given by the mutual information between the secret key and the observations $I(\KK;\OO^{N_o})$, and the equivocation $h_{e}(N_{o})\triangleq H(\KK|\OO^{N_o})$ determines how this leakage decreases the initial lack of information: $h_{e}(N_{o}) =H(\KK)-I(\KK;\OO^{N_o})$. The equivocation is always a non increasing function.
Three things needs to be known to compute these quantities: the distribution of the keys $p_{\KK}$, the distribution of the host signals $p_{\XX}$ and the embedding equation $e(\cdot)$. With this formulation, a perfect covering is tantamount to $I(\KK;\OO^{N_o})=0$. Yet, for most of the watermarking schemes, the information leakage is not null. If identificability is granted, the equivocation about the secret key decreases down to $0$ ($\KK$ is a discrete r.v.) or $-\infty$ ($\KK$ is a continuous r.v.) as the adversary keeps on observing more data.
This information theoretic framework to assess watermarking security has been applied to popular watermarking schemes such
as additive Spread-Spectrum (SS)~\cite{perez:tifs08security,Comesana05},
or DC-QIM (Distortion Compensated Quantization Index Modulation)~\cite{uvigo:tifs06security,4668384}.

This framework is fruitful to establish
if a watermarking scheme is perfectly secure and, if not, to compare the
information leakage of different systems. Nevertheless, it brings little
information regarding T.~Kalker's basic definition of security, e.g.
the ability of the adversary to have access to the watermarking channel.
Indeed, this methodology only needs $p_{\XX}$, $p_{\KK}$ and $e(\cdot)$ to derive the distribution of the observations and, in the end, the equivocation. The decoding side is not taken into account.
Yet, in practice, the estimation of the secret key is only an intermediate goal for the adversary.
The equivocation above defined can be linked to the accuracy of this estimation. 
However, very few works studied the impact of the estimation accuracy on the ability of an unauthorized access to the watermarking channel.

\subsection{Our proposal}\label{sub:SecAlt}
If we look at symmetric cryptography, the security is in direct relationship with the length of the secret key.
The key length $\ell$ in bits defines the number of possible
secret keys as binary words of $\ell$ bits.
The key length provides the maximum
number of tests in logarithmic scale of the brute force attack which finds the
key by scanning the $|\K|$ potential keys~\cite{menezes1997handbook}.
The stopping condition has little importance.
One often assumes that the adversary tests keys until decoded messages are meaningful.
We can also rephrase this with probability:
If the adversary draws a key
uniformly, the probability to pick the secret key is $P=2^{-\ell}$,
 or in logarithmic scale $-\log_{2}(P)=\ell$ bits. 
With the help of some observations, the goal of
the cryptanalysts is to find attacks requiring less operations
than the brute force attack. A good cryptosystem
has a security close to their key length and observing cipher texts is almost useless.
For instance, the best attack so far
on one version of the Advanced Encryption Standard using 128 bits secret key offers
a computational complexity of $2^{126.1}$~\cite{bogdanov2011biclique}. 
Studying security within a probabilistic framework has also been done in other fields of cryptography (for instance, in authentication~\cite{Maurer:2000fk}).

Our idea is to transpose the notion of key length to watermarking.
A crude try is to take the size of the seed of the pseudo-random generator as it is the maximum number of tests of a brute force attack scanning all the seeds. Yet, it doesn't take into account how the secret key is derived from the seed. Another though would be to take the dimension of the space $\K$, but again, it does not consider how watermarking uses the secret key. 
We think that the best approach relies on a probabilistic framework and on the fact that, in watermarking, 
the secret key may not be unique in some sense.
Denote by $\hat{m}$ the message decoded from $\y$ with the secret key $\k$: $\hat{m}=d(\y,\k)$.
We expect that $\hat{m}=m$, but this might be the case for another decoding key $\k'$.
This raises the concept of equivalent keys: for instance, $\k'$ is equivalent to the secret key $\k$ if it grants the decoding of almost all contents watermarked with $\k$. This idea was first mentioned in~\cite{cox2006watermarking}, where the authors made the first distinction between the key lengths in cryptography and watermarking.
The fact that the decoding key might not be unique creates a big distinction with cryptography. However, the rationale of the brute force attack still holds. The attacker proposes a test key $\k'$ and we assume there is a genie telling him whether $\k'$ is equivalent to $\k$. In other words, the security of a scheme does not rely on the difficulty of knowing whether $\k'$ is an equivalent key, but on the rarity of such keys: The lower the probability $P$ of $\k'$ being equivalent to $\k$, the more secure is the scheme.  
We propose to define the effective key length as a logarithmic measure of this probability. Note that in our proposal, we must pay attention to the decoding algorithm $d(\cdot)$ because it is central to the definition of equivalent keys.

Like in the previous methodology, the attack setup (WOA, KMA, KOA) determines the data from which the test key is derived.
In this paper, we restrict our attention to the
Known Message Attack (KMA - an observation is a pair of a watermarked
content and the embedded message: $\OO_{i}=\{\y_{i},m_{i}\}$). 

Assessing the security of watermarking within a probabilistic framework is not new.
S.~Katzenbeisser has also listed the drawbacks of the information theoretic past approach~\cite{Katzenbeisser2005:Computational}.
He especially outlined  the lack of assumption on the computing power of the attacker. 
He then proposed to gauge security as the advantage of the attacker. In a first step, the adversary, modeled by a probabilistic polynomial-time Turing machine, observes contents watermarked with the secret key $\k_{1}$ or $\k_{2}$. Then, the designer produces a new piece of content $\y$ and challenges the adversary whether $\y$ has been watermarked with key $\k_{1}$ or $\k_{2}$. The advantage is defined as the probability of a right guess minus $1/2$. One clearly sees that a strictly positive advantage implies that the adversary has been able to infer some information about the secret key during the first step. However, the relationship with its ability to access the watermarking channel is not straightforward: the decoding is not considered, and the notion of equivalent keys is missing. 

\section{Definition of the effective key length\label{sec:Definition-of-the} }

This section explains the concept of equivalent keys necessary to define the effective key length.
We define by $\D_{m}(k)\subset\X$ the decoding region associated to the message $m$ and for the
key $\k$ by: 
\begin{equation}
\D_{m}(\k)\triangleq\{\y\in\X:d(\y,\k)=m\}.
\end{equation}
The topology and location of this region in $\X$ depends of the
decoding algorithm and of $\k$. 

\begin{figure}[h]
\begin{centering}
\includegraphics[width=0.6\columnwidth]{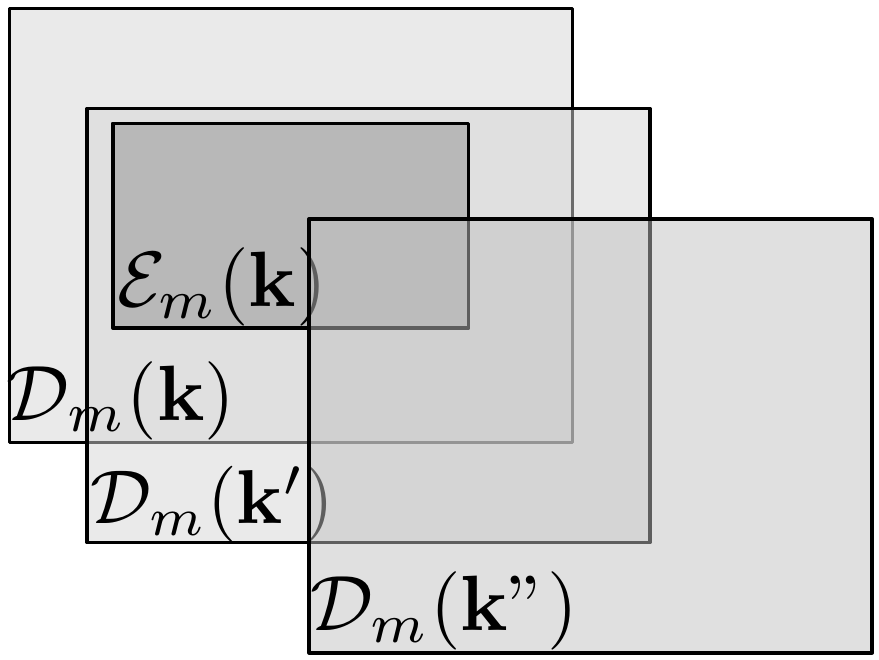}
\par\end{centering}
\caption{Graphical representation in space $\X$ of three decoding regions $\D_{m}(\k)$, $\D_{m}(\k')$ and $\D_{m}(\k'')$ and the embedding region $\E_{m}(\k,0)$: the key $\k'$ belongs the equivalent decoding region $\K_{eq}^{(d)}(\k,0)$ which is not the case for $\k''$.}
\label{figDefkeyY}
\end{figure}

To hide message $m$, the encoder pushes the host vector $\x$ deep
inside $\D_{m}(\k)$, and this creates an embedding region $\E_{m}(\k)\subseteq\X$:
\begin{equation}
\E_{m}(\k)\triangleq\{\y\in\X:\exists\x\in\X \text{ s.t. } \y=e(\x,m,\k)\}.
\end{equation}
A watermarking scheme provides robustness by
embedding in such a way that the watermarked contents are located
far away from the boundary of the decoding region. If the vector extracted
from an attacked content $\z=\y+\n$ goes out of $\E_{m}(\k)$, $\z$
might still be in $\D_{m}(\k)$ and the correct message is decoded.
For some watermarking schemes (like QIM), we have $\E_{m}(\k)\subseteq\D_{m}(\k)$.
Therefore, there might exist another key $\k'$ such that $\E_{m}(\k')\subseteq\D_{m}(\k)$.
A graphical illustration of this phenomenon is depicted on Fig.~\ref{figDefkeyY}.
However, in general even if there is no noise, $\E_{m}(\k)\not\subset\D_{m}(\k)$, and we define the Symbol Error Rate ($\SER$) in the noiseless case as $\eta(0)\triangleq\Prob{d(e(\XX,M,\k),\k)\neq M}$.
Capital letters $\XX$ and $M$ explicit the fact that the probability is over two r.v.: the host and the message to be embedded. 

We now define the equivalent keys and the
associated equivalent region. We make the distinction between
the equivalent decoding keys (the equivalent decoding region) and
the equivalent embedding keys (resp. the equivalent embedding region).

The set of equivalent decoding keys $\K_{eq}^{(d)}(\k,\epsilon)\subset\K$ with $0\leq\epsilon$
is the set of keys that allows a decoding of the hidden
messages embedded with $\k$ with a probability bigger than $1-\epsilon$: 
\begin{equation}
\K_{eq}^{(d)}(\k,\epsilon)=\{\k'\in\K:\Prob{d(e(\XX,M,\k),\k')\neq M}\leq\epsilon\}.
\end{equation}
In the same way, the set of equivalent encoding keys $\K_{eq}^{(e)}(\k,\epsilon)\subset\K$
is the set of keys that allow to embed messages which are reliably
decoded with key $\k$:
\begin{equation}
\K_{eq}^{(e)}(\k,\epsilon)=\{\k'\in\K:\Prob{d(e(\XX,M,\k'),\k)\neq M}\leq\epsilon\}.
\end{equation}

These sets are not empty for $\epsilon\geq\eta(0)$ since $\k$ is then an element.
One expects that, for a sound design, these sets are empty for $\epsilon<\eta(0)$.
Note that for $\epsilon=0$, these two definitions are equivalent
to:
\begin{equation}
\K_{eq}^{(d)}(\k,0)=\{\k'\in\K:\E_{m}(\k')\subseteq\D_{m}(\k)\},
\end{equation}
and
\begin{equation}
\K_{eq}^{(e)}(\k,0)=\{\k'\in\K:\E_{m}(\k)\subseteq\D_{m}(\k')\}.
\end{equation}

\begin{figure}[h]
\begin{centering}
\includegraphics[width=0.7\columnwidth]{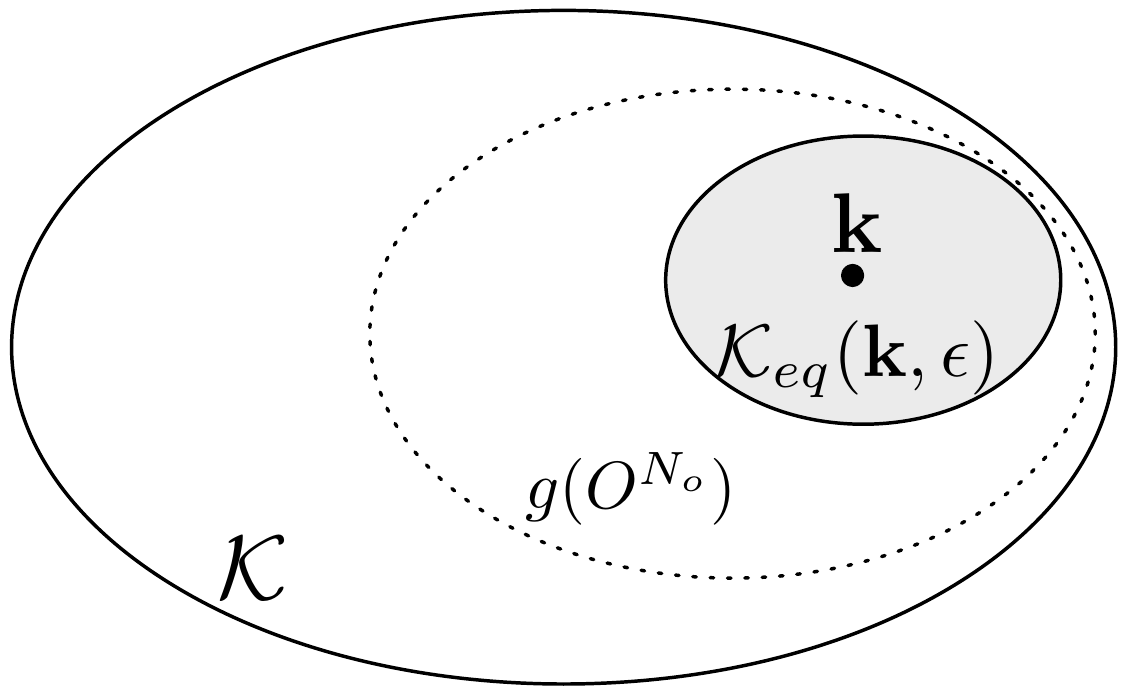}
\par\end{centering}
\caption{Graphical representation of the key space $\K$ and the equivalent region $\K_{eq}(\k)$. The dotted boundary represents the support of the generative function $g(O^{N_{o}})$ which is used to draw new keys when the adversary get observations.}
\label{figDefKeyK}
\end{figure}

The effective key length of a watermarking scheme is now defined using theses definitions. The adversary draws a key $\k'\in\K$ taking into account the set of observations $\OO^{N_{o}}$ with a generative function $\KK'=g(\OO^{N_{o}})$. The function $g(\cdot)$ is either deterministic or stochastic (such that $\KK'\sim p(\k|\OO^{N_{o}})$ for instance). A graphical example of the key space $\K$ and the equivalent region $\K_{eq}(\k)$ is depicted on Fig.~\ref{figDefKeyK} together with the support region of a potential generative function.

The probability $P^{(d)}(\epsilon,N_{o})$ (or $P^{(e)}(\epsilon,N_{o})$) that the adversary
picks up a key belonging to the equivalent decoding region (resp. equivalent
embedding region) is:
\begin{equation}
\begin{array}{c}
P^{(d)}(\epsilon,N_{o})=
\Exp{\KK}{\Exp{\OO^{N_{o}}}{\Exp{\KK'}{\KK'\in\K_{eq}^{(d)}(\KK,\epsilon)|\OO^{N_{o}}}}},
\end{array}
\end{equation}
and similarly for $P^{(e)}(\epsilon,N_{o})$.
Finally, by analogy with cryptography, the effective key length translates this probability into bits as follows: 
\begin{equation}
\ell^{(d)}(\epsilon,N_{o})\triangleq-\log_{2}(P^{(d)}(\epsilon,N_{o}))\quad\text{bits},\label{eq:keyLd}
\end{equation}
and similarly for $\ell^{(e)}(\epsilon,N_{o})$. 
Note also that for some watermarking schemes, we have $\K_{eq}^{(e)}(\k,\epsilon)=\K_{eq}^{(d)}(\k,\epsilon)$. There is then no need to make a distinction and we will denote the probability and the effective key length as $P(\epsilon,N_{o})$ and $\ell(\epsilon,N_{o})$. Additionally, we call $\ell(\epsilon,0)$ the {\it basic key length}, i.e. the effective key length of a watermarking system when no observation is available.  

We conclude this section by stating that the size of the seed is the maximum value of the effective  key length. We assume that the pseudo-random generator is public (Kerckhoff's principle) so that nothing prevents the attacker from using this generator. If any different two seeds produce two different secret keys, then a brute force attack on the seed yields a key length of the size of the seed. Nevertheless, the attacker may work with a different pseudo-random generator. The theoretical study below assumes that he uses a perfectly random generator giving $\KK'\sim p_{\KK}$ for $N_{o}=0$, or that he uses $\KK'=g(\OO^{N_{o}})$ for $N_{o}>0$.
In practice, the value of the effective key length should be clipped to the size of the seed in bits.

\section{Theoretical effective key length computations\label{sec:Theoretical-computations}}
The goal of this section is to compute the expressions of the key
length for the most popular class of watermarking schemes: additive spread-spectrum.

\subsection{The equivalent region}
Consider a spread spectrum one-bit watermarking s.t. $\y=e(\x,m,\k)=\x+(-1)^{m}\alpha\k$,
with $m\in\{0,1\}$. The host is modeled by a white Gaussian vector of size $N_v$
and power $\sigma_{X}^{2}$. The secret key is a pseudo-random unitary vector 
($\|\mathbf{k}\|=1$) and $\K$ is consequently the unit hyper-sphere. The parameter $\alpha$ controls the Document to Watermark power Ratio with the following relation:
\begin{equation}\label{eq:alpha}
\alpha=\sqrt{N_{v}}\sigma_{X}10^{-\frac{\DWR}{20}}.
\end{equation} 

The decoder is correlation based: $d(\y,\k)=0$ if $\y^{\top}\k>0$, $1$ else.
We assume that $\y$ is corrupted by an independent white Gaussian noise of power $\sigma_{N}^{2}$. The $\SER$ is given by
\begin{equation}\label{eq:eta}
\eta(\sigma_{N})=\Phi\left(-\frac{\alpha}{\sqrt{\sigma_{X}^{2}+\sigma_{N}^{2}}}\right) 
\end{equation}
with $\Phi(\cdot)$ the cumulative distribution function of the standard normal random variable. Eq.~\eqref{eq:alpha} and~\eqref{eq:eta} show that the robustness of the scheme quantified by $\eta(\sigma_{N})$ is an increasing function of $N_{v}$.

The adversary uses the same encoding or decoding functions but with a different key $\k'$ with $\|\k'\|=1$. We restrict our attention to the equivalent decoding keys. The reason is that $\K_{eq}^{(d)}(\k,\epsilon)=\K_{eq}^{(e)}(\k,\epsilon)$ because $d(e(\x,m,\k'),\k)$ and $d(e(\x,m,\k),\k')$ have identical pdfs. We define by $\theta$ the angle between $\k$ and $\k'$: $\cos\theta=\k^{\top}\k'$. 
The adversary's decoding statistic is $\y^{\top}\k'\sim\mathcal{N}((-1)^{m}\alpha\cos\theta,\sigma_{X}^{2})$ and his $\SER$ is 
\begin{equation}
\epsilon=\Phi\left(\frac{-\alpha\cos\theta}{\sigma_{X}}\right).
\end{equation}
For a given $\epsilon\geq\eta(0)$, $\k'$ is an equivalent key if its angle with $\k$ is lower than 
\begin{eqnarray}
\theta_{\epsilon}&\triangleq&\arccos{(-\Phi^{-1}(\epsilon)\sigma_{X}/\alpha)}\\
&=&\arccos{\left(-\frac{\Phi^{-1}(\epsilon)}{\sqrt{N_{v}}}10^{\frac{\DWR}{20}}\right)}.\label{eq:theta_epsilon}
\end{eqnarray}
$\K_{eq}(\epsilon,\k)$ is the intersection of the unit hypersphere and the single inner hypercone of axis $\k$ and angle $\theta_{\epsilon}$, i.e. a spherical cap.

\subsection{The basic key length}\label{subsec:lengthSS=0}
For $N_{o}=0$, the probability that a key $\k'$ uniformly distributed over $\K$ is 
inside $\K_{eq}(\epsilon,\k)$ is the ratio of the solid angle of
this spherical cap and the full hypersphere (see Appendix~\ref{Sec:Fdistribution}): 
\begin{equation} 
\label{eq:cone}
P_{SS}(\epsilon,0)=\frac{1-I_{\cos^{2}(\theta_{\epsilon})}(1/2,(N_{v}-1)/2)}{2},
\end{equation}
where $I(\cdot)$ is the regularized incomplete beta function.
Fig.~\ref{fig:pgfKeySSNo0} shows that, contrary to $\eta(\sigma_{N})$, the basic key length is a decreasing function of $N_{v}$ for fixed $\epsilon$ and $\DWR$.  This illustrates 
the trade-off between security and robustness. Appendix~\ref{Sec:Fdistribution} gives the asymptotical value of the basic key length:
\begin{equation}\label{eq:asym}
\lim_{N_{v}\rightarrow\infty} P_{SS}(\epsilon,0) = \frac{1}{2}\left(1-\text{erf}\left(\frac{|\Phi^{-1}(\epsilon)|}{\sqrt{2}}10^{\frac{\DWR}{20}}\right)\right).
\end{equation}
This means that SS schemes become more robust as $N_{v}\rightarrow\infty$ but their basic key length does not vanish to 0.  
 
\subsection{Key length for $N_{o}>0$\label{subsec:lengthSS>0}}
For $N_{o}>0$, we suppose without loss of generality that the embedded
messages were all set to $0$ (if not, we work with $(-1)^{m_{i}}.\y_{i}$). One possible estimator $\hat{\k}$ is to compute the average of $\{\y_{i}\}_{i=1}^{N_{o}}$ and to normalize it.
The probability of this estimation being inside $\K_{eq}(\epsilon,\k)$
is approximated by the cumulative distribution function of a non-central
F-distribution variable of degrees of freedom $\nu_{1}=1$, $\nu_{2}=N_{v}-1$
and noncentrality parameter $\lambda=\alpha^{2}\frac{N_{o}}{\sigma_{X}^{2}}$,
weighted by the probability $\Prob{\k'^{\top}\k>0}$ (see Appendix~\ref{Sec:Fdistribution}): 
\begin{eqnarray}
\label{eq:gamma}
P_{SS}(\epsilon,N_{o})\approx&\left[1-F\left(\frac{(N_{v}-1)\cos^{2}(\theta_{\epsilon})}{1-\cos^{2}(\theta_{\epsilon})};1,N_{v}-1,\lambda\right)\right]\nonumber\\
&*\Phi\left(\sqrt{\lambda}\right).
\end{eqnarray}
The experimental work below shows that this approximation is indeed very accurately in our setup.

\section{Practical effective key length computations\label{sec:Practical-computations}}

Depending of the watermarking scheme, the effective key length defined by (\ref{eq:keyLd}) may not have a literal formula and this section aims at giving an experimental setup for its estimation. We first propose a general framework with a high complexity. For the case of additive spread spectrum, some simplifications occur and stems into a more practical experimental setup.

\subsection{The general framework \label{subsec:Monte-carlo}}

If we are not limited in term of computational power, the probability $P^{(d)}(\epsilon,N_{o})$ can be approximated using a classical Monte-Carlo method. We first generate a set of $N_1$ random secret keys $\{\mathbf{k}_{i}\}_{i=1}^{N_{1}}$. For each of them, we also generate $N_2$ test keys $\{\mathbf{k}_{i,j}'\}_{j=1}^{N_{2}}$. Then, an estimation is: 
\begin{equation}
\hat{P}^{(d)}(\epsilon,N_{o})=\frac{1}{N_{1}N_2}\sum_{i=1}^{N_{1}}\sum_{j=1}^{N_{2}}u^{(d)}(\mathbf{k}_{i,j}',\epsilon),
\label{eq:keyMonteCarlo}
\end{equation}
 where 
\begin{equation}
\begin{array}{cccc}
u^{(d)}(\mathbf{k}_{i,j}',\epsilon) & =1 & \mathrm{if} & \mathbf{k}'_{i,j}\in\mathcal{K}^{(d)}_{eq}(\mathbf{k}_{i},\epsilon)\\
 & =0 & \mathrm{else.}
\end{array}
\end{equation}
The probability $P^{(e)}(\epsilon,N_{o})$ is respectively approximated using the indicator function $u^{(e)}(\cdot)$ of $\mathcal{K}^{(e)}$.

For $N_{o}=0$, each test key $\k'_{i,j}$ is independently drawn according to $p_{\KK}$. For $N_{o}>0$, we first generate a set of $N_{o}$ observations $\OO_{i}^{N_{o}}$ depending on $\mathbf{k}_{i}$,  and we resort to a specific estimator to construct $\mathbf{k}'_{i,j}=g(\OO_{i}^{N_{o}})$ (see Sec.~\ref{sec:Definition-of-the}).

Secondly, the equivalent region may not have a defined indicator function. 
In this case, we generate $N_t$  other contents $\{\y_{\ell}\}_{\ell=1}^{N_{t}}$ watermarked with $\k_{i}$ (resp. original contents) and the test is satisfied if at least $(1-\epsilon)N_t$ contents are correctly decoded (respectively embedded) using $\k'_{i,j}$. Mathematically, for the decoding equivalence:
\begin{equation}\label{eq:counting}
\mathbf{k}'_{i,j}\in\mathcal{K}^{(d)}_{eq}(\mathbf{k}_{i},\epsilon) \approx |\{\y_{\ell}\in\D_{m_{\ell}}(\k'_{i,j})\}|>(1-\epsilon)N_{t}.
\end{equation}
In this case an estimation of $P^{(d)}(\epsilon,N_{o})$ needs $N_{1}(N_{2}N_{o}+N_{t})$ embeddings and $N_{1}N_{2}N_{t}$ decodings. Due to the limitation of the Monte-Carlo method, $N_{1}N_{2}$ should be in the order of $1/P^{(d)}(\epsilon,N_{o})$ for having a meaningful relative variance of the estimation. The parameter $N_{t}$ should also be quite big for having a good approximation of the indicator function of $\mathcal{K}^{(d)}_{eq}(\mathbf{k}_{i},\epsilon)$. It is reasonable to take $N_{t}=O(c^{N_{v}})$ for some constant $c$ where $N_{v}$ is the dimension of the space $\X$ containing $\D_{m_{\ell}}(\k'_{i,j})$.

This procedure is generic and it blindly resorts to the embedding and the decoding algorithms as black boxes. If we have some knowledge about the watermarking technique, some tricks reduce the complexity of the estimation. First, the probability of finding an equivalent key might not depend on $\k_{i}$, so that we can restrict to $N_{1}=1$ original key. This is the case for spread spectrum technique. For $N_{o}=0$, the probability to be estimated may be very weak and out of reach of the Monte-Carlo method. We can use rare event probability estimator such as the one proposed in~\cite{CEROU:2011:INRIA-00584352:1}. Last but not least, for a given $\k'_{i,j}$, the geometry of $\D_{m}(\k'_{i,j})$ can help reducing $N_{t}$ and still obtaining a good approximation of the indicator function of $\mathcal{K}^{(d)}_{eq}(\mathbf{k}_{i},\epsilon)$. The following subsections put into practice these simplifications for the additive spread spectrum technique.

\subsection{Approximation of the equivalent region $\mathcal{K}^{(d)}_{eq}$}\label{sub:Approximate}

The equivalent region $\mathcal{K}^{(d)}_{eq}$ depends on the embedding and decoding. For the additive spread spectrum, both processes are so simple that we were able to derive closed-form formula of the probability in Sect.~\ref{sec:Theoretical-computations}. We suppose now that the embedding is more complex which prevents theoretical derivations.
We will pretend in Sect.~\ref{sub:Experiments} that the Improved Spread Spectrum proposed in~\cite{Malvar03} plays the role of such an embedding.

For a given host $\x$, we can always express the result of the embedding as
\begin{equation}
\y = e(\x,m,\k) = a(\x,m)\k + b(\x,m)\u_{\bot}(\x,m),
\end{equation}
where $\k^{\top}\u_{\bot}(\x,m)=0$. The decoding with $\k'$ is based on the quantity:
\begin{equation}
\y^{\top}\k' = a(\x,m)\cos(\theta)+b(\x,m).(\k'^{\top}\u_{\bot}(\x,m)),
\end{equation}
whose sign yields the decoded bit $\hat{m}$. It is important to note that the decoding step using a test key $\k'$ can be performed in a 2 dimensional space spanned by $(\k,\u_{\bot}(\x,m))$.
The Symbol Error Rate is expressed in term of the CDF of the statistical r.v. $\YY^{\top}\k'$  which depends on $\theta$, and is thus denoted $\SER(\theta)$.
For $\theta=0$, we have $\SER(0)=\eta(0)$. For $\epsilon\geq\eta(0)$, we define
\begin{equation}\label{eq:thetaExp}
\theta_{\epsilon}=\max_{\SER(\theta)=\epsilon}\theta.
\end{equation}
This shows that the equivalent decoding region is a hypercone of axis $\k$ and angle $\theta_{\epsilon}$ which depends on the embedding. The only thing we need is to experimentally estimate angle $\theta_{\epsilon}$. Then, we use Eq.~\eqref{eq:cone} in order to obtain an approximation of the effective key length.

\begin{figure}[h]
  \centering \includegraphics[width=1\columnwidth]{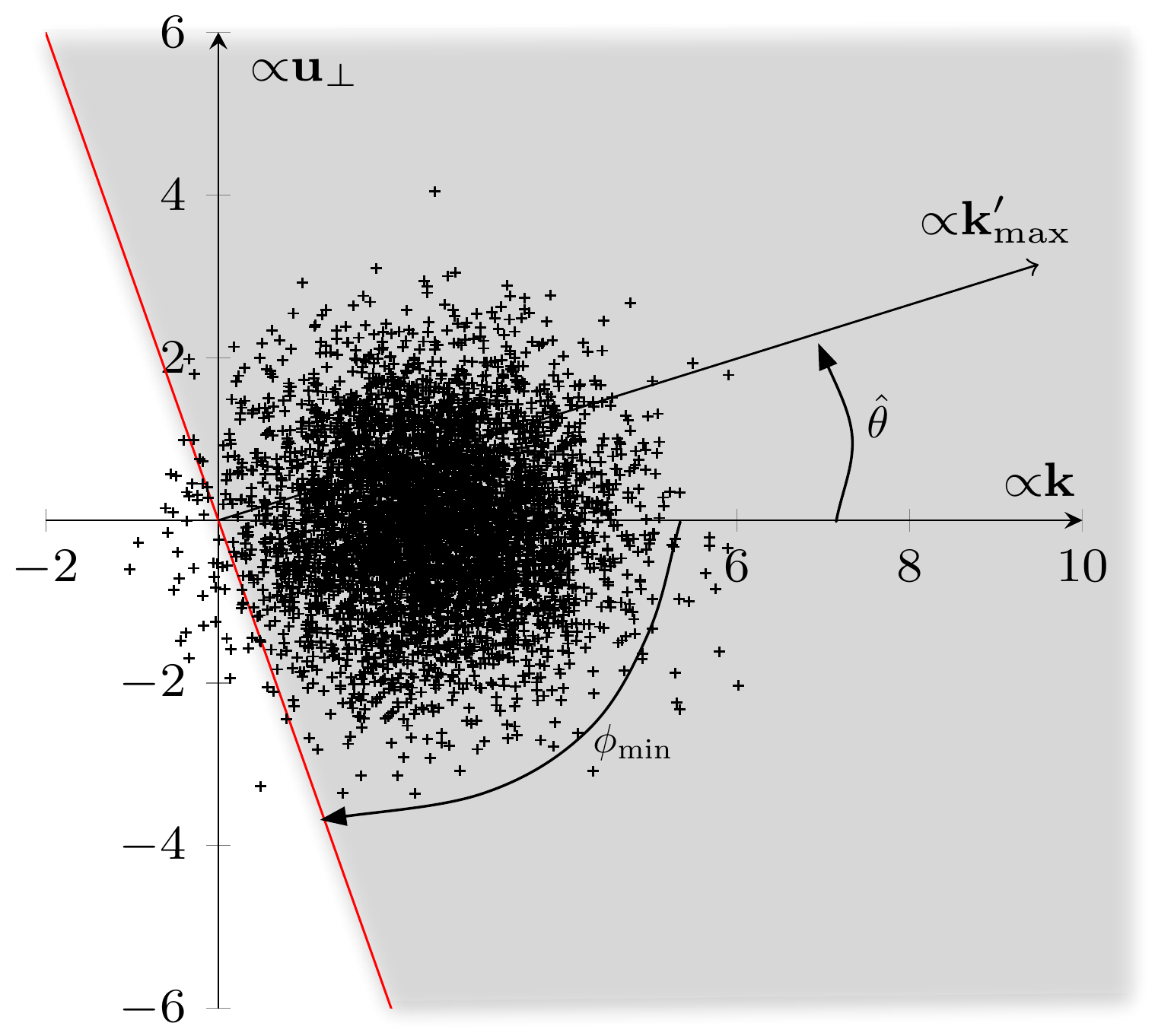}
  \caption{Projections of $N_t=5000$ watermarked vectors ($N_v=60$, $m=1$) on $\k$ and $\mathbf{u}_{\perp}$, $\DWR=10$\,dB, $N_v=60$, $\epsilon=10^{-2}$. The vector $\k'_\mathrm{max}$ correctly decodes $[(1-\epsilon)N_t]$ contents.}
  \label{fig:pgfScatter-crop.pdf} 
\end{figure}

The estimation of $\theta_{\epsilon}$ is made under the following rationale. A vector $\y$ watermarked by $\k$ with $m=1$ is correctly decoded by any $\k'$ s.t. $\k'^{\top}\k\geq\cos(\theta_{\epsilon})$ if its angle $\phi$ with $\k$ is such that $\phi \in [\theta_{\epsilon}-\pi/2,\theta_{\epsilon}+\pi/2]$ (see Fig. \ref{fig:pgfScatter-crop.pdf}). 
In practice, we generate $N_t$ contents $\{\y_i\}_{i=1}^{N_{t}}$ watermarked with $m=1$, and we compute their angles $\{\phi_{i}\}_{i=1}^{N_{t}}$ with $\k$. Once sorted in increasing order, we iteratively find the angle $\phi_{\min}$ such that $\mathrm{int}((1-\epsilon)N_t)$ vectors have their angle $\phi \in [\phi_{\min}-\pi/2,\phi_{\min}+\pi/2]$ and set $\hat{\theta}_{\epsilon}=\pi/2-\phi_{\min}$. 

A much lower number of vectors is needed to accurately estimate one parameter than a full region of the space.
$N_t$ and $N_v$ directly impact the accuracy of $\hat{\theta}_\epsilon$, but since this boils down to the estimation of a single parameter, the magnitude of $N_t$ is rather low in comparison with the effective key length. For example, at $N_v=60$ and $\DWR=10$\,dB, we generate $N_t=10^6$ contents in order to obtain a reliable effective key length of more than 100 bits, whereas an estimation based on~\eqref{eq:counting}  would have required $N_t \approx 2^\ell \times 10^3 \approx 10^{33}$ contents. Moreover, the angle $\theta_{\epsilon}$ is the same for any $\k$, so the estimation is done only once. This avoids the counting of correct decodings over $N_{t}$ vectors of~\eqref{eq:counting}.

\subsection{Rare event probability estimator}\label{sub:RareEvent} 

A fast rare event probability estimator\footnote{available as a Matlab toolbox at~\url{www.irisa.fr/texmex/people/furon/src.html}} is described in~\cite{Guyader2011}. We explain its application for the case $N_{o}=0$.
This algorithm estimates the probability $\Prob{s(\KK')>0}$ under $\KK'\sim p_{\KK}$. It needs three ingredients: the generation of test keys distributed according to $p_{\KK}$, the distribution invariant modification of test keys, and the soft score function $s(\cdot)$.

We work with an auxiliary random vector $\WW\sim\mathcal{N}(\mathbf{0},\mathbf{I}_{N_{v}})$. The generator draws $\WW$ and outputs a test key $\KK' = \WW/\|\WW\|$. Since the distribution of $\WW$ is isotropic, $\KK'$ is uniformly distributed over the hypersphere. The algorithm draws $n$ such test keys, and iteratively modifies those having a low score. The modification takes back $\WW$,  adds an independent noise $\NN\sim\mathcal{N}(\mathbf{0},\mathbf{I}_{N_{v}})$, and scales the result: $\WW'=(\WW+\mu\NN)/\sqrt{1+\mu^{2}}$. Parameter $\mu$ controls the strength of the modification. In the end, it returns a new test key $\WW'/\|\WW'\|$. For any value of $\mu$, the modification lets the distribution invariant because $\WW'\sim\mathcal{N}(\mathbf{0},\mathbf{I}_{N_{v}})$. The properties of this algorithm depends on $n$ as given in~\cite{Guyader2011}. Qualitatively, the bigger $n$ is, the more accurate but slower is this estimator.

We propose two score functions depending on whether we know the equivalent region $\mathcal{K}^{(d)}_{eq}$:
\subsubsection{$\K_{eq}^{(d)}$ is known (Sect.~\ref{sec:Theoretical-computations}) or approximated (Sect.~\ref{sub:Approximate})\label{subsubREEst}}
the score function is simply a metric between the test key and the border of the equivalent region: $s(\KK') = \KK'^{\top}\k - \cos(\hat{\theta}_{\epsilon})$. In the end, the algorithm returns an estimation of $\Prob{\cos(\theta)>\cos(\hat{\theta}_{\epsilon})}$ when $\KK'$ is uniformly distributed over the hypersphere. 

\subsubsection{$\K_{eq}^{(d)}$ is not known\label{subsubREGen}} 
We generate $N_t$ contents $\{\y_{i}\}_{i=1}^{N_{t}}$ watermarked with $\k$, and the score function is the $\mathrm{int}(\epsilon N_t)$-th smallest `distance' from these vectors to the set $\D_{m}(\k')$, where $\mathrm{int}(.)$ denotes the closest integer function. For SS or ISS, this `distance' is for instance the correlation $\k'^{\top}\y$. In the end, the algorithm returns an estimation that $\mathrm{int}((1-\epsilon)N_t)$ vectors are correctly decoded, when $\KK'$ is uniformly distributed over the hypersphere.

\section{Results and Discussions}\label{sub:Experiments}
The goal of the experimental part is twofold.
First, we wish to assess the soundness of the experimental measurement of the effective key length. This is done by a comparison to the theoretical results for the additive Spread Spectrum. Second, we would like to illustrate the trade-off between security and robustness. Third, we compare the additive Spread Spectrum (SS) to the Improved Spread Spectrum (ISS)~\cite{Malvar03}.

In the latter method, the embedding has two parameters $(\beta,\gamma)$: $e(\x,m,\k)=\x+(-1)^{m}(\beta-\gamma(\x^{\top}\k))\k$. For a fair comparison, the parameters $N_{v}$, $\epsilon$, $\sigma_{X}$ and $\DWR$ are fixed. This implies that
\begin{equation}
\alpha^{2}=\beta^{2}+\gamma^{2}\sigma_{X}^{2}=N_{v}\sigma_{X}^{2}10^{-\frac{\DWR}{10}}.
\end{equation}
The robustness is gauged by using a AWGN channel of variance $\sigma_{N}^2$ giving a Watermark to Noise Ratio $\WNR=10\log_{10}(\sigma_{W}^2/\sigma_{N}^2)$ dB. As for the security, we use $N_{t}=10^{6}$ contents to estimate $\hat{\theta}_{\epsilon}$ for $N_{o}=0$ as explained in Sect.~\ref{sub:Approximate}.
The two embedding functions, SS and ISS, produce different angles. Then, the rare event probability estimator is used as described in Sect.~\ref{sub:RareEvent} with $n=80$. For $N_{o}>0$, the attacker's key estimator $g(\cdot)$ is just the normalized average of vectors $\{(-1)^{m_{i}}\y_{i}\}_{i=1}^{N_{o}}$ as explained in App.~\ref{Sec:Fdistribution}. It appears that the probabilities to be estimated are dramatically bigger, and the Monte Carlo method of Sect.~\ref{subsec:Monte-carlo} is good enough.



\subsection{The impact of embedding parameters $N_v$ and $\DWR$} 
\begin{figure}[h]
\centering \includegraphics[width=1\columnwidth]{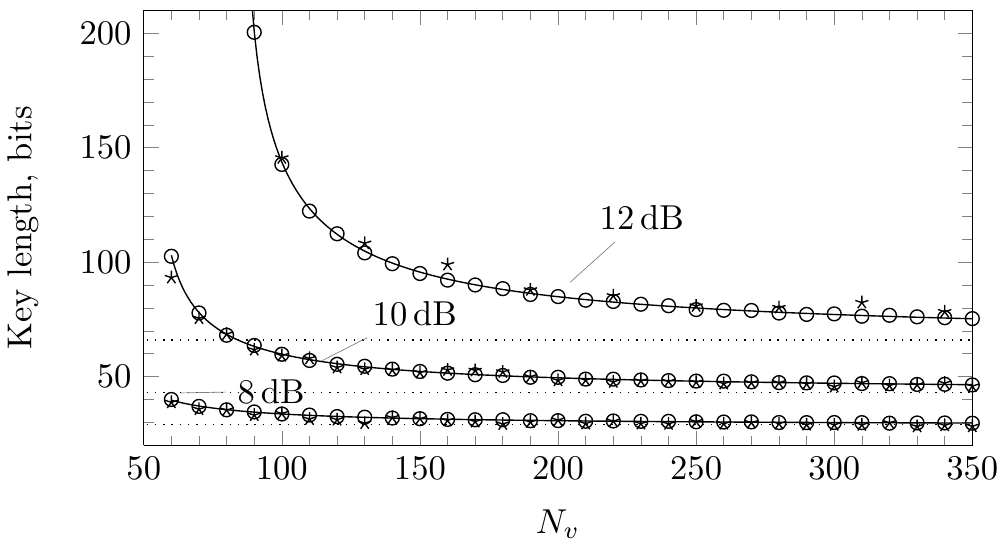}
\caption{The basic key lengths for $\epsilon=10^{-2}$ and $\DWR\in\{8, 10, 12\}$ using the theoretical expression (\ref{eq:cone}) (plain lines), estimation of the equivalent region presented in Sect.~\ref{sub:Approximate} with $N_t=10^6$ ($\circ$) and rare event analysis presented in Sect.~\ref{subsubREGen} ($\star$) with $N_t=5.10^4$ and $n=80$. The horizontal dotted lines are the asymptotes~(\ref{eq:asym}).}
\label{fig:pgfKeySSNo0} 
\end{figure}

\begin{figure}[h]
\centering \includegraphics[width=1\columnwidth]{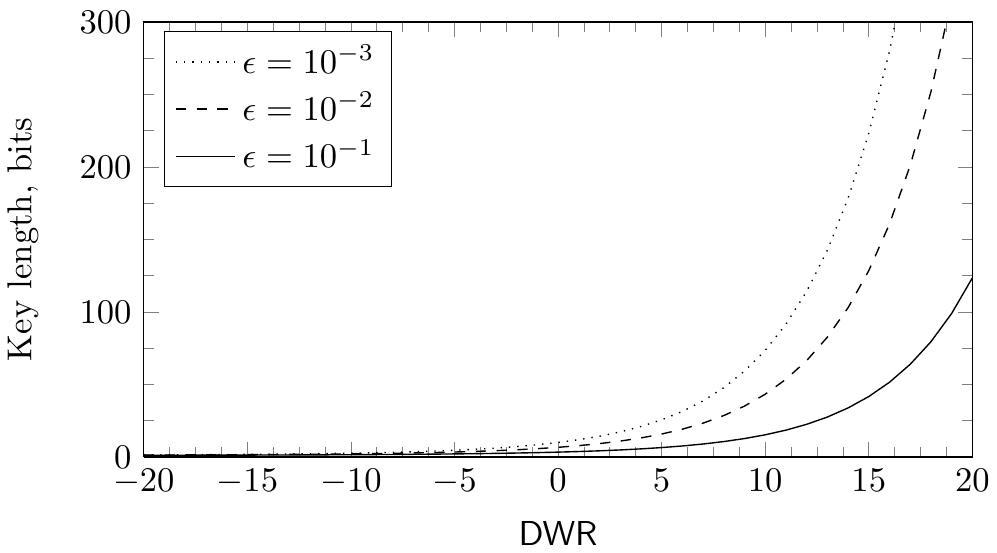}
\caption{Basic key length for hosts of infinite length given in~(\ref{eq:asym}).}
\label{fig:pgfKeySSAsympt} 
\end{figure}

Fig.~\ref{fig:pgfKeySSNo0} points out the decrease of the basic key length w.r.t. $N_{v}$ for a constant embedding distortion. Contrary to a statement of~\cite[Sec. 4.1]{cox2006watermarking}, the effective key length is not proportional to $N_{v}$. We can also note the relatively fast convergence to the strictly positive asymptote~\eqref{eq:asym}, especially at high embedding distortions. 
Fig. \ref{fig:pgfKeySSAsympt} highlights the decrease of this asymptotic key length with the embedding distortion. The basic key length is computationally significant, say above 64 bits, only for $\DWR$ greater than 12 dB for $\epsilon=0.01$.
If the watermarking technique is such that a lower $\DWR$ remains imperceptible, it should not be recommended from a security point of view.

\subsection{The impact of security parameters $\epsilon$ and $N_o$}
The decrease of the basic key length with $\epsilon$ is confirmed on Fig. \ref{fig:pgfKeySSAsympt}. This is not a surprise: the more stringent the access to the watermarking channel, the higher the security is.

\begin{figure}[h]
\centering \includegraphics[width=1\columnwidth]{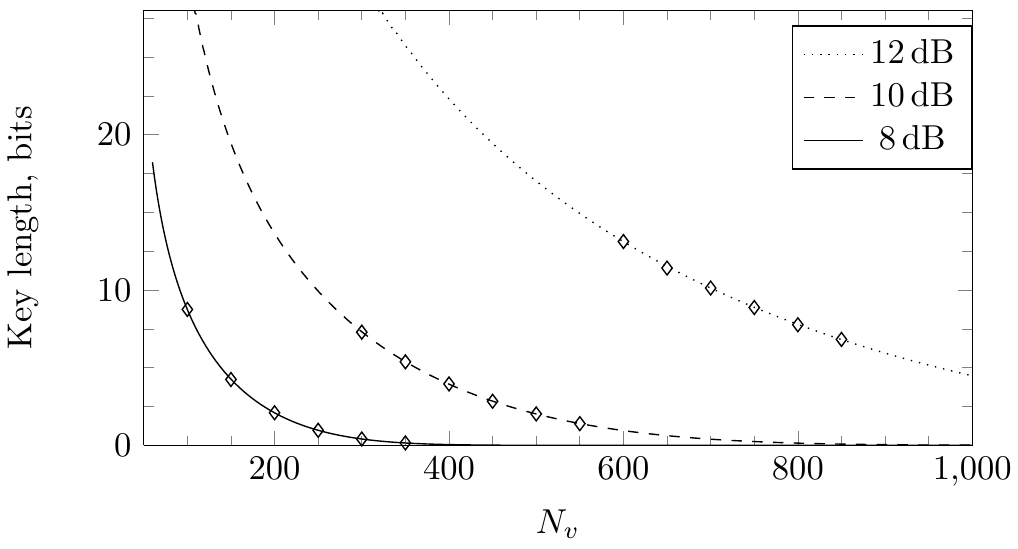}
\caption{Key lengths for $\epsilon=10^{-2}$, $N_o=1$, and different $\DWR$ using approximation~\eqref{eq:gamma} and Monte-Carlo simulations of Sect.~\ref{subsec:Monte-carlo} ($\diamond$) with $N_1=1$ and $N_2=10^6$.}
\label{fig:pgfKeySSNo1} 
\end{figure}

\begin{figure}[h]
\centering \includegraphics[width=1\columnwidth]{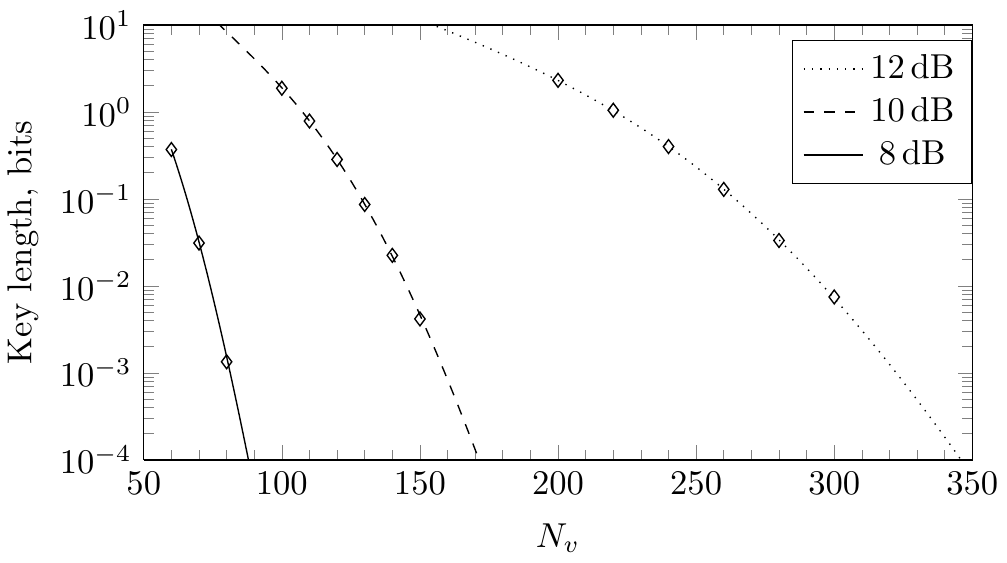}
\caption{Key lengths for $\epsilon=10^{-2}$, $N_o=10$, and different $\DWR$ using approximation~\eqref{eq:gamma} and Monte-Carlo simulations of Sect.~\ref{subsec:Monte-carlo} ($\diamond$) with $N_1=1$ and $N_2=10^6$.}
\label{fig:pgfKeySSNo10} 
\end{figure}

Fig.~\ref{fig:pgfKeySSNo1} and Fig.~\ref{fig:pgfKeySSNo10} illustrate the dramatical decrease of the effective key length when observations are available in the KMA context. For example,  at $\DWR=10$\,dB, $N_v=300$ and $\epsilon=10^{-2}$, the effective key length drops from roughly 50 bits to 8 bits for $N_o=1$ and nearly 0 bits for 10 observations. In brief, SS watermarking is not secure if the attacker gets observations. Note also that the approximation~\eqref{eq:gamma} is very close to the Monte Carlo estimations.

\subsubsection{The interplay between security and robustness}
\begin{figure}[h]
\centering \includegraphics[width=1\columnwidth]{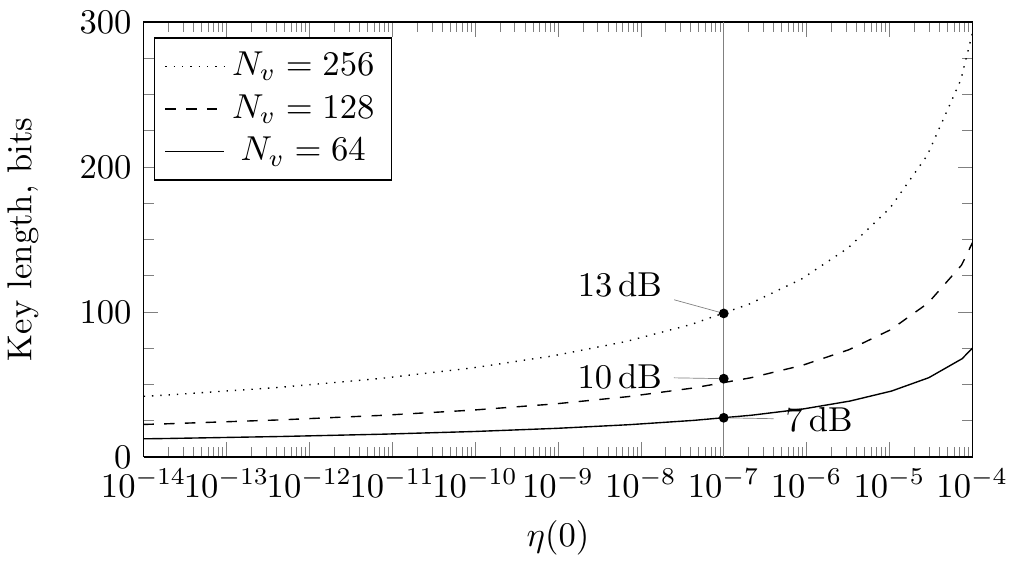}
\caption{Trade-off between robustness and security. The plot is computed by varying $\DWR$ ; the ticks show the values of $\DWR$ for $\eta(0)=10^{-7}$.}
\label{fig:pgfSecVsRobNv} 
\end{figure}

Fig.~\ref{fig:pgfSecVsRobNv} shows the trade-off between robustness  measured by $\eta(0)$ and security gauged by $\ell$.
For a given robustness, the longer the host, the better the security and the smaller the distortion of the scheme.
Conversely, to decrease $\eta(0)$ while keeping the basic key length constant, it is better to increase $N_v$ than to increase the distortion. This is due to the fact that the effective key length decreases to a strictly positive value w.r.t. $N_v$ but on the other hand decreases to zero w.r.t. the embedding distortion. Fig.~\ref{fig:pgfSecVsRobNv} highlights that $\ell$ and $N_{v}$ both decrease w.r.t the distortion at a constant robustness, as already suggested by Fig.~\ref{fig:pgfKeySSAsympt}. 


\begin{figure}[h]
\centering \includegraphics[width=1\columnwidth]{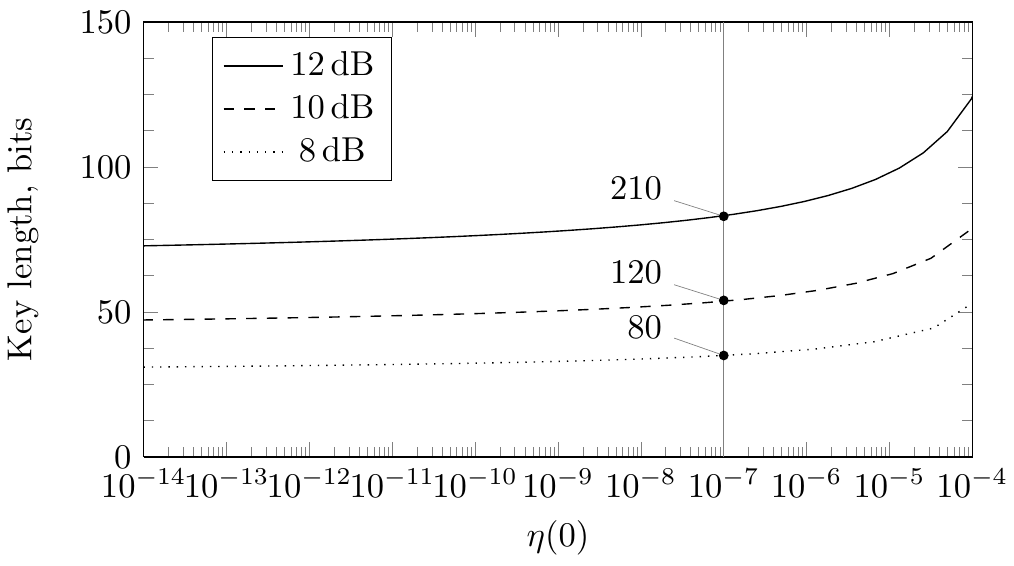}
\caption{Trade-off between robustness and security. The plot is computed by varying $N_v$ ; the ticks show the values of $N_v$ for $\eta=10^{-7}$.}
\label{fig:pgfSecVsRobDWR} 
\end{figure}

\begin{figure}[h]
\centering
\begin{tabular}{c}
\includegraphics[width=1\columnwidth]{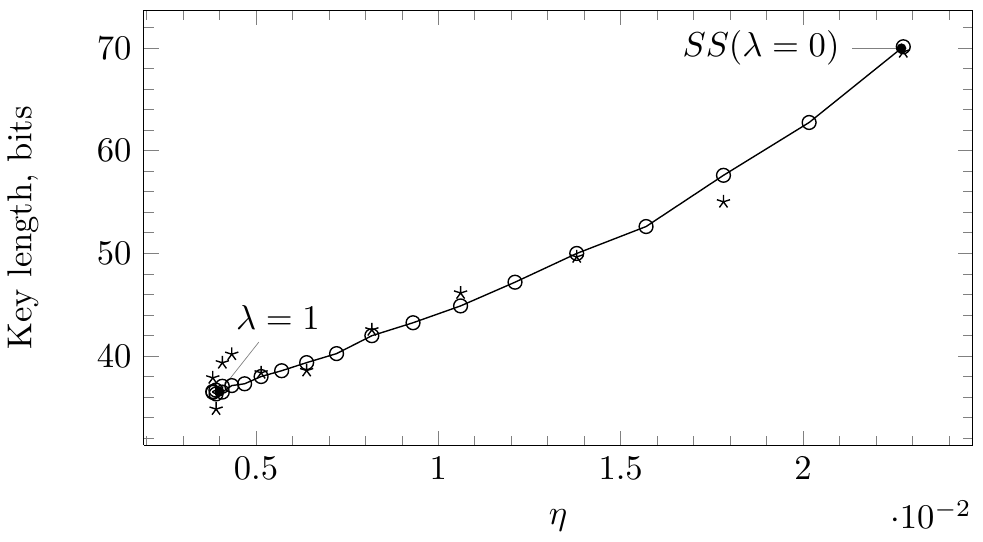}\\
$\WNR= -10\,\mathrm{dB}$\\
\includegraphics[width=1\columnwidth]{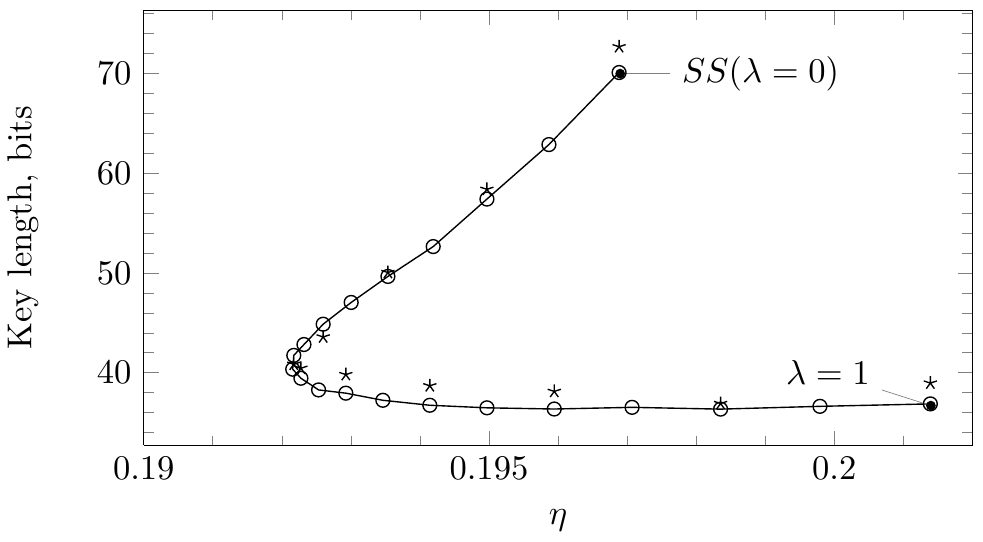}\\
$\WNR= -20\,\mathrm{dB}$
\end{tabular}
\caption{Trade-off between robustness and security for ISS. The plot is computed by varying $\lambda$ at $\DWR= 10$ dB, $N_v=80$, and $\epsilon=10^{-2}$. The key length is estimated using the method of Sect.~\ref{sub:Approximate} ($o$) with $N_t=10^6$ and the rare events estimator of Sect.~\ref{subsubREGen} with $N_t=5.10^4$ and $n=80$ ($\star$).}
\label{fig:pgfEtaVsKeyISS-cropWNR-10Nv80DWR10.pdf} 
\end{figure}

We now compare SS with ISS regarding both security and robustness. Fig.~\ref{fig:pgfEtaVsKeyISS-cropWNR-10Nv80DWR10.pdf} shows that the host rejection parameter $\lambda$ has a negative impact on the key length and a mitigated positive impact on the robustness. At low $\WNR$ regimes, two different $\lambda$ may give the same robustness but two different effective key lengths. One should consequently choose the $\lambda$ parameter maximizing the security in this case.

\subsubsection{The validity of the practical approaches}
The practical methods (Monte-Carlo, rare-event estimator or equivalent region estimation) match the literal formula (\ref{eq:cone}) and (\ref{eq:gamma}) either for small or large effective key lengths on Figures~\ref{fig:pgfKeySSNo0}, \ref{fig:pgfKeySSNo1} and \ref{fig:pgfKeySSNo10}. The rare event estimator (Sect.~\ref{subsubREGen}) and the estimator based on $\hat{\theta}_{\epsilon}$ (Sect.~\ref{sub:Approximate}) are particularly accurate for large key lengths (see Figures~\ref{fig:pgfKeySSNo0} and \ref{fig:pgfEtaVsKeyISS-cropWNR-10Nv80DWR10.pdf}), whereas the Monte-Carlo estimator is more efficient for small key length (see Figures~\ref{fig:pgfKeySSNo1} and \ref{fig:pgfKeySSNo10}).

\section{Conclusion\label{sec:Conclusion}}
In this paper, we have proposed a new measure called the effective key length to characterize watermarking security.
Contrary to symmetric cryptography, there are several keys granting to access to the watermarking channel. This gives birth to the notion of equivalent keys. The effective key length represents the difficulty of finding such a key.

We have computed the effective key length theoretically and practically for additive spread spectrum schemes. The main conclusions of this specific analysis are the following. For a constant error rate against the AWGN channel, the effective key length increases w.r.t. the length of the host and decreases w.r.t. the distortion. Contrary to what was stated in~\cite{cox2006watermarking}, the effective key length is not proportional to the size of the host. The decrease of the effective key length is dramatic regarding the number of observations in the KMA context, which strongly supports the idea of changing the embedding key as frequently as possible.

Our future work will apply this methodology to other watermarking schemes (such as DC-QIM) but also to other scenario attacks such as the Oracle attack.

\appendices
\section{Probabilities for Spread Spectrum}
\label{Sec:Fdistribution}
Let $\XX\sim\mathcal{N}(\mu\mathbf{e}_{1},\sigma^{2}\mathbf{I}_{N_{v}})$, where $\mathbf{e}_{1}$ is the first vector of the canonical basis of $\mathbb{R}^{N_{v}}$. The appendix gives the probability that the normalized correlation $D = \XX^{\top}\mathbf{e}_{1} / \|\XX\|$ is above a threshold $\tau$. A simpler problem is the computation of:
\begin{eqnarray}
\Prob{D^{2}>\tau^{2}}&=&\Prob{\frac{X_{1}^{2}}{\sum_{i=1}^{N_{v}}X_{i}^{2}}>\tau^{2}}\\
&=&\Prob{\frac{X_{1}^{2}}{\sum_{i=2}^{N_{v}}X_{i}^{2}}>\frac{\tau^{2}}{1-\tau^{2}}}\\
&=&\Prob{\frac{X_{1}^{2}}{(N_{v}-1)^{-1}\sum_{i=2}^{N_{v}}X_{i}^{2}}>\frac{(N_{v}-1)\tau^{2}}{1-\tau^{2}}}\label{eq:F}
\end{eqnarray}
Denote $F=\frac{X_{1}^{2}}{(N_{v}-1)^{-1}\sum_{i=2}^{N_{v}}X_{i}^{2}}$. For $\mu=0$, $F$ is the ratio of two independent $\chi^{2}$ random variables of degree of freedom $\nu_{1}=1$ and $\nu_{2}=N_{v}-1$, thus it is distributed as a Snedecor F-distribution $F(1,N_{v}-1)$~\cite[26.6]{Abramowitz1964:Handbook}, whose CDF is given by a regularized incomplete beta function $I_{\frac{x}{x+N_{v}-1}}(1/2,(N_{v}-1)/2)$, and
\begin{equation}
\Prob{D^{2}>\tau^{2}} = 1-I_{\tau^{2}}(1/2,(N_{v}-1)/2).
\end{equation}
This is the probability that a centered white Gaussian vector lies inside a two-nappe hypercone of angle $\arccos(\tau)$. By symmetry around the origin, we have for the single nappe hypercone $\Prob{D>\tau} = (1-I_{\tau^{2}}(1/2,(N_{v}-1)/2))/2$. This holds indeed for any random vector $\XX$ whose distribution is symmetric wrt to the origin, and in particular for a uniform distribution over the hypersphere. This proves~\eqref{eq:cone} if one sets $\k=\mathbf{e}_{1}$ and $\tau=\cos(\theta_{\epsilon})$. Another point is that as $N_{v}\rightarrow\infty$, the distribution of $F$ converges to a $\chi^{2}_{1}$ distribution~\cite[26.6.11]{Abramowitz1964:Handbook} while the RHS of the inequality in~\eqref{eq:F} converges to $\kappa^{2}$ if $\tau=\kappa/\sqrt{N_{v}}$. Therefore, $\lim_{N_{v}\rightarrow\infty}\Prob{D>\tau} = (1-\text{erf}(|\kappa|/\sqrt{2}))/2$. This proves~\eqref{eq:asym} because $\cos\theta_{\epsilon}=\kappa/\sqrt{N_{v}}$ due to~\eqref{eq:theta_epsilon}.

For $\mu>0$, $F$ has a non-central F-distribution with noncentrality parameter $\lambda=\mu^{2}/\sigma^{2}$ and degrees of freedom $\nu_{1}=1$ and $\nu_{2}=N_{v}-1$, whose CDF is denoted by $F(x;1,N_{v}-1,\lambda)$. Therefore,
\begin{equation}
\Prob{D^{2}>\tau^{2}} = 1-F\left(\frac{(N_{v}-1)\tau^{2}}{1-\tau^{2}};1,N_{v}-1,\lambda\right).
\end{equation}
However, the argument of symmetry no longer holds for deriving $\Prob{D>\tau}$. We propose to write:
\begin{eqnarray}
\Prob{D>\tau}&=&\Prob{(D^{2}>\tau^{2})\&(D>0)}\\
&=&\Prob{D^{2}>\tau^{2}|D>0}.\Prob{D>0}\\
&\approx&\Prob{D^{2}>\tau^{2}}.\Prob{D>0},
\end{eqnarray}
with $\Prob{D>0} = \Phi(\sqrt{\lambda})$. This approximation is accurate for $\lambda\rightarrow0$ and $\lambda\rightarrow+\infty$.

The link with Spread-Spectrum for $N_{o}>0$ is the following. The attacker estimates the secret key as $\hat{\KK}=\bar{\YY}/\|\bar{\YY}\|$ with $\bar{\YY}$ the average of the observations:
\begin{equation}
\bar{\YY} = \frac{1}{N_{o}}\sum_{i=1}^{No}\YY_{i}=\alpha \k + \frac{1}{N_{o}}\sum_{i=1}^{No}\XX_{i}=\alpha \k + \bar{\XX}.
\end{equation}
If we assume that the hosts are independent white Gaussian vectors, then $\bar{\XX}\sim\mathcal{N}(\mathbf{0},\frac{\sigma_{X}^{2}}{N_{o}}\mathbf{I})$. Now, $\hat{\KK}$ is an equivalent key  (\textit{ie.} it belongs to the spherical cap) iff $\bar{\YY}$ belongs to the inner single-nappe hypercone: $D=\k^{\top}\bar{\YY}/\|\bar{\YY}\|\geq \cos(\theta_{\epsilon})$, which translates into
\begin{equation}
D=\frac{U_{1}+\sqrt{\lambda}}{\sqrt{\sum_{i=1}^{N_{v}}U_{i}^{2}}}\geq\tau = \cos(\theta_{\epsilon}),
\end{equation}
where $\UU=(U_{1},\cdots,U_{N_{v}})$ is the projection of $\bar{\XX}$ on a basis of $\mathbb{R}^{N_{v}}$, whose first vector is $\k$, divided by $\sigma_{X}^{2}/N_{o}$ so that $\UU\sim\mathcal{N}(0,\mathbf{I})$. After the transformation that turns $D$ into the r.v. $F$, it appears that this latter has a noncentral F-distribution with a noncentrality parameter
\begin{equation}\label{eq:lambda}
\lambda=\alpha^{2}N_{o}/\sigma_{X}^{2}=N_{v}N_{o}.10^{-\frac{\DWR}{10}}.
\end{equation}
This provides the approximation~\eqref{eq:gamma} in the text.
  
In the same way as above, $F$ converges to a non central $\chi^{2}_{1}$  modelled as $(U_{1}+\sqrt{\lambda})^{2}$ when $N_{v}\rightarrow\infty$. This makes $\Prob{D^{2}>\tau^{2}}\rightarrow\Prob{(U_{1}+\sqrt{\lambda})^{2}>\kappa^{2}}$.
Parameter $\lambda$ linearly increases with $N_{v}$ as shown in~\eqref{eq:lambda}. Inspired by~\cite[Proof of Lemma~2.1]{Christian1990101}, we write:
\begin{eqnarray*}
\Prob{(U_{1}+\sqrt{\lambda})^{2}>\kappa^{2}}&=&\Prob{U_{1}^{2}+\lambda^{2}+2U_{1}\sqrt{\lambda}>\kappa^{2}}\\
&=&\Prob{U_{1}>-\frac{1}{2}\sqrt{\lambda}+\frac{\kappa^{2}-Y^{2}}{2\sqrt{\lambda}}}\stackrel{\lambda\rightarrow\infty}{\rightarrow} 1 
\end{eqnarray*}
and so does $\Phi(\sqrt{\lambda})$. In the end, $\lim_{N_{v\rightarrow\infty}}\Prob{D>\tau}=1$
which shows that the effective key length vanishes to zero as $N_{v}\rightarrow\infty$ provided that $N_{o}>0$. 


\bibliographystyle{IEEEtran}


\end{document}